\documentclass[fleqn,usenatbib]{mnras}

\usepackage{newtxtext,newtxmath}

\usepackage[T1]{fontenc}

\DeclareRobustCommand{\VAN}[3]{#2}
\let\VANthebibliography\thebibliography
\def\thebibliography{\DeclareRobustCommand{\VAN}[3]{##3}\VANthebibliography}

\usepackage{graphicx}	
\usepackage{amsmath}	
\usepackage{tabularx}
\usepackage{subcaption}
\usepackage{physics}
\usepackage{siunitx}
\usepackage{float}

\title[LOFAR imaging of lensed quasars]{Gravitational lensing in LoTSS DR2 -- Extremely faint 144-MHz radio emission from two highly magnified quasars}

\author[J. P. McKean et al.]{J. P. McKean,$^{1,2}$\thanks{E-mail: mckean@astron.nl}
R. Luichies,$^2$ 
A. Drabent,$^3$
G. G\"urkan,$^3$
P. Hartley,$^4$
A. Lafontaine,$^1$
I. Prandoni,$^5$
\newauthor H. J. A. R\"ottgering,$^6$ 
T. W. Shimwell,$^{1,6}$
H. R. Stacey$^{7}$
and  C. Tasse$^{8,9}$
\\
$^1$ASTRON, Netherlands Institute for Radio Astronomy, Oude Hoogeveensedijk 4, 7991 PD, Dwingeloo, the Netherlands\\
$^2$Kapteyn Astronomical Institute, University of Groningen, P.O. Box 800, 9700AV Groningen, the Netherlands\\
$^3$Th\"uringer Landessternwarte, Sternwarte 5, D-07778 Tautenburg, Germany\\
$^4$Square Kilometre Array Organisation, Macclesfield, Cheshire SK11 9DL, United Kingdom\\
$^5$INAF-Istituto di radioastronomia, Via P. Gobetti 101, Bologna, I-40129, Italy\\
$^6$Leiden Observatory, Leiden University, PO Box 9513, NL-2300 RA Leiden, the Netherlands\\
$^7$Max Planck Institute for Astrophysics, Karl-Schwarzschild Str. 1, D-85748 Garching bei M\"unchen, Germany\\
$^8$GEPI \& USN, Observatoire de Paris, CNRS, Universit\'e Paris Diderot,    5 place Jules Janssen, 92190 Meudon, France\\
$^9$Centre for Radio Astronomy Techniques and Technologies, Department of Physics and Electronics, Rhodes University, Grahamstown 6140, South Africa
}

\date{Accepted 2021 March 29. Received 2021 March 28; in original form 2021 January 8}

\pubyear{2021}

\begin{document}
\label{firstpage}
\pagerange{\pageref{firstpage}--\pageref{lastpage}}
\maketitle

\begin{abstract}
We report extremely faint 144 MHz radio emission from two gravitationally lensed quasars, SDSS~J1004+4112 ($z = 1.730$) and SDSS~J2222+2745 ($z = 2.803$), using the LOFAR Two Metre Sky Survey (LoTSS) data release 2. After correcting for the lensing magnifications, the two objects have intrinsic flux-densities of $13\pm2$ and $58\pm6$~$\mu$Jy, respectively, corresponding to 144 MHz rest-frame luminosities of $10^{23.2\pm0.2}$ and $10^{24.42\pm0.05}$~W~Hz$^{-1}$, respectively. In the case of SDSS~J1004+4112, the intrinsic flux density is close to the confusion limit of LoTSS, making this radio source the faintest to be detected thus far at low frequencies, and the lowest luminosity known at $z\gtrsim0.65$. Under the assumption that all of the radio emission is due to star-formation processes, the quasar host galaxies are predicted to have star-formation rates of $5.5^{+1.8}_{-1.4}$ and $73^{+34}_{-22}$~M$_{\odot}$~yr$^{-1}$, respectively. Further multi-wavelength observations at higher angular resolution will be needed to determine if any of the detected radio emission is due to weak jets associated with the quasars.
\end{abstract}

\begin{keywords}
quasars: individual: SDSS~J1004+4112 -- quasars: individual: SDSS~J2222+2745 -- gravitational lensing: strong
\end{keywords}



\section{Introduction}
\label{introduction}

Hydrodynamical simulations predict that present-day massive elliptical galaxies grew hierarchically, with a large fraction of their stellar mass being formed through violent mergers in the early Universe (e.g. \citealt{schaye2015,vogelsberger2014}). In this scenario, the rapid onset of star-formation ($>10^2$~M$_\odot$~yr$^{-1}$) is fuelled through significant gas accretion and collapse during the merger phase, producing a starburst that is extremely luminous at rest-frame far-infrared (FIR; $L_{\rm IR}>10^{12}$~L$_{\odot}$) wavelengths (e.g. \citealt{sanders1988}). Also, mergers can provide a mechanism to push gas onto the central black hole of the galaxy, leading to significant accretion and black hole growth, and the eventual triggering of an active galactic nucleus (AGN; e.g. \citealt*{dimateo2005}). Radiative and mechanical feedback associated with star formation and AGN activity is then expected to clear the host galaxy of dust and gas, revealing the previously obscured quasar and preventing further significant star formation and black hole growth (e.g. \citealt{hopkins2005}). Although further merger events over cosmic time may result in a new starburst/AGN phase, such galaxies are expected to passively evolve until present day, with only small amounts of star formation produced through minor mergers and gas accretion.

This evolutionary model successfully explains the high levels of dust-obscured star formation in quasar host galaxies at redshifts between 1 and 4, where galaxy growth, star formation and AGN activity peak (e.g. \citealt*{madau2014}), and there is substantial evidence that mergers and feedback play an important role in galaxy evolution at these epochs (e.g. \citealt{heckman2014,hodge2020}). In particular, resolved observations at mm-wavelengths of optically/X-ray luminous quasars detect large molecular gas reservoirs and heated dust that is consistent with extreme levels of star formation ($>10^3$~M$_\odot$~yr$^{-1}$), close to Eddington-limited star-formation rate intensities, and evidence of nearby companions in around 30~percent of cases \citep{beelen2006,coppin2008,banerji2017,nguyen2020,stacey2020}. However, testing the full evolutionary path of this model requires identifying quasar hosts at high redshift that have lower levels of star formation, even though there is significant molecular gas still available. At high redshifts, this is only possible using the increased sensitivity and angular-resolution provided by gravitational lensing. For example, a study of 104 (mainly optically selected) lensed quasars with the {\it Herschel Space Observatory} detected heated dust in 66 percent of the sources, with cold dust temperatures (median of $38^{+12}_{-5}$~K) and a range of FIR luminosities that were consistent with significant levels of star-formation (median of $120^{+160}_{-80}$~M$_{\odot}$~yr$^{-1}$; \citealt{stacey2018}).

Studying lensed quasar hosts at high redshift with even lower star-formation rates ($<40$~M$_{\odot}$~yr$^{-1}$) is more challenging, given the sensitivity and source confusion of FIR survey telescopes. In this respect, wide-field surveys at radio wavelengths offer the possibility to identify quasar hosts that have low levels of star formation via the radio--infrared correlation (e.g. \citealt{calistro2017,gurkan2018,wang2019}). The Low Frequency Array (LOFAR; \citealt{vanhaarlem2013}) is currently undertaking the LOFAR Two Metre Sky Survey (LoTSS; \citealt{shimwell2019}), which aims to image the northern sky at 144 MHz with an angular resolution of 6 arcsec and an rms of 70~$\mu$Jy~beam$^{-1}$ at optimal declinations. From the first data release, covering 454~deg$^2$ of the sky, three lensed quasars were found to have 144 MHz radio emission that was consistent with star formation at the level of 60 to 210~M$_{\odot}$~yr$^{-1}$, given their correlated radio and FIR luminosities \citep{stacey2019}. The LoTSS data release two (DR2; Shimwell et al., in prep.) covers a sky area of 5720 deg$^2$ that includes 63 known lensed quasars in the survey footprint, of which around two-thirds are detected ($>3\sigma$) at 144 MHz (Luichies et al., in prep). The vast majority of the detections are consistent with star formation at a similar level to that found with the (albeit smaller) DR1 sample of lensed quasars \citep{stacey2019}, although a number of objects also have significant low frequency emission that is almost certainly associated with radio jets (Badole et al., in prep.; Lafontaine et al., in prep.). 

In this letter, we focus on two of the lensed quasars from the LoTSS-DR2, namely SDSS~J1004+4112 ($z = 1.730$) and SDSS~J2222+2745 ($z = 2.805$), which stand out from the rest of the sample due to their very high total lensing magnifications (67 and 32, respectively). As a result, multiple lensed images from the same object are resolved by LOFAR for the first time, and extremely faint radio emission is detected, which in one case is close to the LoTSS confusion limit. In Section \ref{data}, we briefly summarise the LOFAR and archival radio data for both lensed quasars, and in Section~\ref{discussion} we discuss the nature of their 144 MHz radio emission and the future prospects of using LOFAR and gravitational lensing to probe the faint radio source population at high redshift. Throughout, we assume a flat $\Lambda$CDM Universe with $H_0 = \SI{67.8}{km.s^{-1}.Mpc^{-1}}$, $\Omega_{\mathrm{M}} = 0.31$ and $\Omega_{\Lambda} = 0.69$ \citep{planck2016}. The radio spectral index $\alpha$ is defined as $S_{\nu} \propto \nu^{\alpha}$, where $S_{\nu}$ is the flux density and $\nu$ is the frequency.

\section{Observations \& Results}
\label{data}

\subsection{LoTSS 144 MHz imaging}

SDSS~J1004+4112 and SDSS~J2222+2745 were observed at a central frequency of 144 MHz with the High Band Array (HBA) of LOFAR as part of LoTSS on 2015 August 25 and 2019 October 3, respectively. Both datasets were acquired following the standard LoTSS observing procedure, that is, an 8 h observation consisting of 230 frequency sub-bands between 120 and 167 MHz, each with a bandwidth of 195 kHz (Shimwell et al., in prep.). The standard LoTSS calibration and imaging pipelines were used to produce the final mosaics of the target fields, which had a restoring beam of 6 arcsec \citep{shimwell2019,tasse2020}. An inspection of the mosaics found no significant imaging artefacts in the regions of sky containing the two lensed quasars; therefore, no further calibration and imaging was needed. The off-source rms map noise was 74 and 104~$\mu$Jy~beam$^{-1}$ for SDSS J1004+4112 and SDSS J2222+2745, respectively.

\begin{figure*}
    \centering
    \includegraphics[width=18cm]{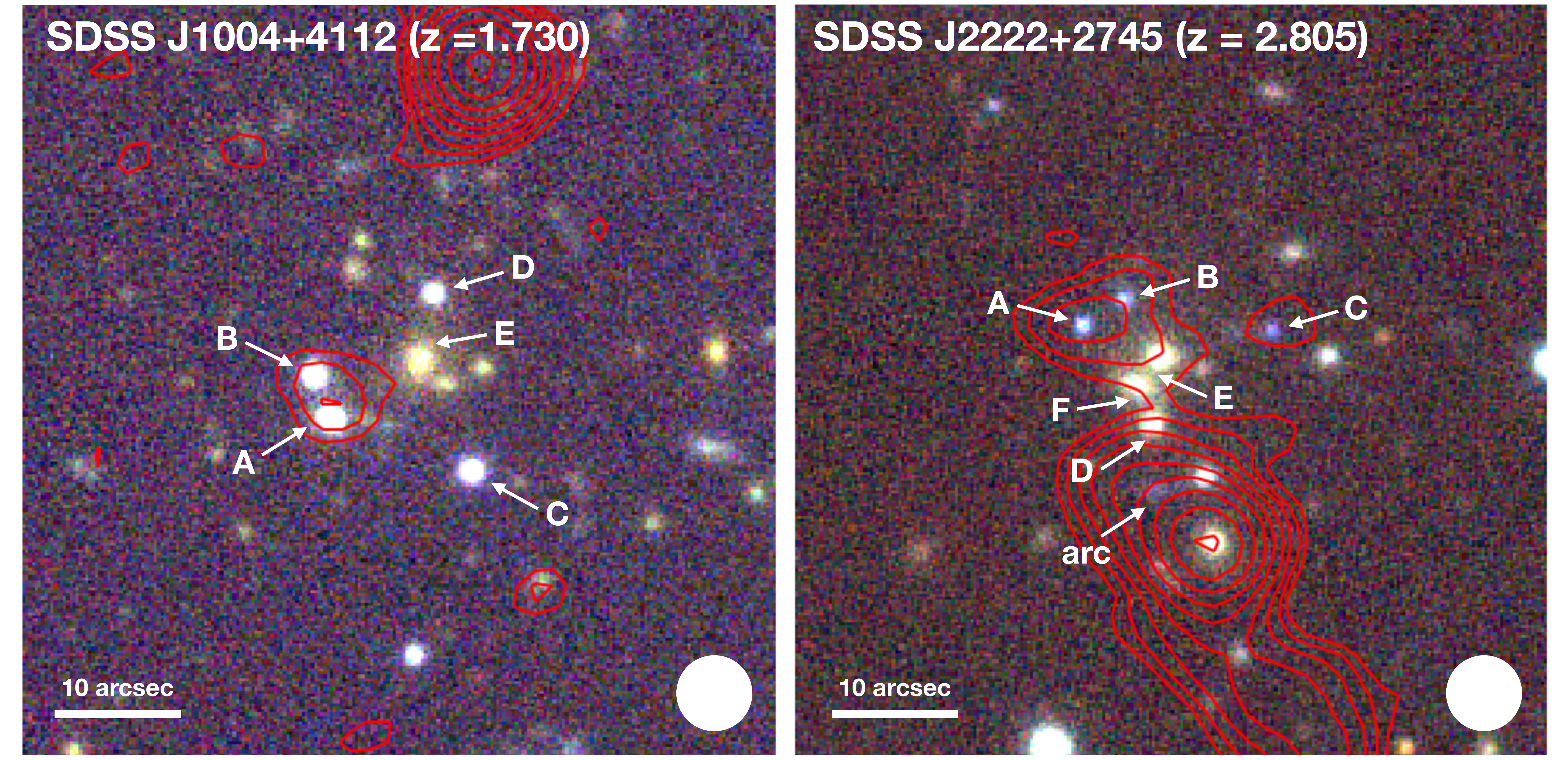}
    \caption{Pan-STARRS $r$, $i$ and $z$ combined colour image with the LOFAR 144 MHz emission contours shown in red for SDSS J1004+4112 (left) and SDSS J2222+2745 (right). The LOFAR surface brightness contours are at $\sigma_{\rm rms} \times (-3, 3, 4.24, 6, 8.49, 12,...)$ where $\sigma_{\rm rms}$ is the rms map noise, which is 74 and 104~$\mu$Jy~beam$^{-1}$ for SDSS J1004+4112 and SDSS J2222+2745, respectively. The 6 arcsec restoring beam size is shown as the filled circle in the bottom right corner of each plot. There is clear radio emission associated with lensed images A and B from SDSS J1004+4112, and lensed images A, B and C from SDSS J2222+2745. The non-detection of emission from lensed images C, D and E from SDSS J1004+4112 is consistent with the expected magnification of those images and the sensitivity of the LOFAR imaging. The optical (and possible radio) emission from lensed images D, E and F of SDSS J2222+2745 is blended with the foreground lensing galaxies, and is not considered further here.}
    \label{fig1}
\end{figure*}

In Fig.~\ref{fig1} the 144 MHz surface brightness contours for each system overlaid on the $r$, $i$ and $z$ colour image from Pan-STARRS is presented. The radio emission from each system was modelled using delta functions to estimate the total flux density from each detected lensed image (see Table~\ref{table1}). This was done in the image plane using the {\sc imfit} task within the Common Astronomy Software Applications package. We discuss the results for each system separately.

\subsection{SDSS~J1004+4112}

SDSS~J1004+4112 is a broad emission line quasar at redshift 1.730 that is gravitationally lensed by an intervening galaxy cluster at redshift 0.68 into five images with a maximum separation of 14.62 arcsec \citep{inada2003}. The positions of the lensed images are in a fold configuration, where the two closest lensed images, A and B, have the highest point-source magnifications of 29.7 and 19.6, respectively, and the three other images, C, D and E have point-source magnifications of 11.6, 5.8 and 0.16, respectively \citep{oguri2010}. Previous radio imaging with the Very Large Array at 5 GHz detected emission from lensed images A, B, C and D, implying an intrinsic (lensing corrected) flux density at 5 GHz of $2.5\pm0.2$~$\mu$Jy \citep{jackson2011}.

The LoTSS image of SDSS~J1004+4112 at 144~MHz detects low frequency radio emission from lensed images A and B at the $6\sigma$-level, with a combined peak surface brightness of 430~$\mu$Jy~beam$^{-1}$. The measured flux-densities of A and B at 144 MHz are $369\pm76$ and $271\pm75$~$\mu$Jy, respectively, which given their lensing magnifications implies an intrinsic (lensing corrected) flux density of $13\pm2$~$\mu$Jy. This intrinsic flux density is about 27 times lower than the $5\sigma$ detection threshold of LoTSS, and is close to the LOFAR 6 arcsec beam-size confusion limit ($\sim8$ to 11~$\mu$Jy for a point source; \citealt{sabater2020}). SDSS~J1004+4112 is currently the faintest radio source detected by LOFAR. No 144 MHz radio emission was detected from lensed images C, D and E, which is to be expected given the $3\sigma$ upper-limit to their flux densities and their lower lensing magnifications.

The radio spectral index between 144 MHz and 5 GHz is found to be $\alpha_{\rm 144~MHz}^{\rm 5~GHz}=-0.46\pm0.05$, under the assumption that this is described by a single power-law model, which is rather flat when compared to a typical starburst galaxy (see Section \ref{discussion} for further discussion). Using this spectral index, the rest-frame 144 MHz luminosity density is $L_{\rm 144~MHz} = 1.58\pm0.08 \times10^{23}$~W~Hz$^{-1}$, making SDSS~J1004+4112 the lowest radio-power source currently known at $z\gtrsim0.65$ that has been detected with LOFAR (e.g. \citealt{kondapally2020,sabater2020}). Under the assumption that all of this radio emission is associated with star formation, and using the redshift-dependent radio--infrared correlation at 150 MHz derived by \citet{calistro2017}, we find that this luminosity density is equivalent to a radio-derived star-formation rate of $5.5^{+1.8}_{-1.4}$~M$_{\odot}$~yr$^{-1}$.

\begin{table}
\centering
\caption{Summary of the emission from SDSS J1004+4112 and SDSS J2222+2745, including the 5 GHz flux density ($S_{\rm 5~GHz}$) from \citet{jackson2011} and the 144 MHz flux density ($S_{\rm 144~MHz}$) reported here, and the point-source magnification ($\mu$) from lens modelling by \citet{oguri2010} and \citet{sharon2017}, respectively. Non-detections assume point source emission and are quoted at the $3\sigma$ level.}
\label{table1}
\begin{tabular}{llrrrrr} 
Source name		& Lensed image	& $S_{\rm 5~GHz}$	& $S_{\rm 144~MHz}$	& $\mu$ 	\\ 
				&				& ($\mu$Jy) 		& ($\mu$Jy)			&		\\ \hline
SDSS~J1004+4112	& A				& $64\pm8$ 		& $369\pm76$			& 29.7	\\
				& B				& $39\pm8$ 		& $271\pm75$			& 19.6	\\
				& C				& $30\pm8$ 		& $<222$				& 11.6	\\
				& D				& $33\pm8$ 		& $<222$				& 5.8		\\	
				& E				& $<21$			& $<222$				& 0.16	\\
SDSS~J2222+2745	& A				& 				& $749\pm110$			& 14.5      \\
				& B				& 				& $606\pm108$		& 10.8	\\
				& C				& 				& $496\pm107$		& 6.7		\\ \hline
\end{tabular}
\end{table}

\subsection{SDSS~J2222+2745}

SDSS~J2222+2745 is a gravitationally lensed quasar at redshift 2.805 that was identified from follow-up observations of a candidate lensing cluster at redshift 0.49 \citep{dahle2013}. The lensing cluster has three similar-mass elliptical galaxies at the centre, which together produce a complex set of lensing caustics. This results in six lensed images of SDSS~J2222+2745 being formed, with a maximum image separation of 15.1 arcsec. Lensed images A, B and C have the highest point-source magnifications of 14.5, 10.8 and 6.7, respectively, whereas lensed images D, E and F have point-source magnifications close to unity (1.43, 0.76 and 0.95, respectively) due to their proximity to the central galaxies of the lensing cluster \citep{sharon2017}. In addition to the lensed quasar, there is a giant gravitational arc at redshift 2.30 that is highly distorted, but is only singly imaged, and there are three other lensed galaxies that are multiply imaged, one of which is at redshift 4.56. There are no published high frequency radio measurements of SDSS~J2222+2745 that resolve the lensed images. 

The LoTSS image of SDSS~J2222+2745 at 144 MHz is quite complex; there is clear radio emission associated with lensed images A, B and C, but there is also extended radio emission at the position of the lensing galaxies within the cluster. The emission from lensed images A and B is blended with a peak surface brightness of 745~$\mu$Jy~beam$^{-1}$, which is equivalent to a 7$\sigma$ detection. Lensed image C is well resolved from the other lensed images and the lensing galaxies, and is detected at the 4$\sigma$-level; this is the first time that gravitationally lensed images of the same background source have been fully resolved with LOFAR. The flux densities of lensed images A, B and C at 144 MHz are found to be $749\pm110$, $606\pm108$ and $496\pm107$~$\mu$Jy, respectively. Given their lensing magnifications, this implies an intrinsic (lensing corrected) flux density of $58\pm6$~$\mu$Jy for SDSS~J2222+2745. Similar to above, this intrinsic flux density is below the detection threshold of LoTSS. The radio emission from lensed images D, E and F is predicted to be below the rms noise level of the data. Therefore, the radio emission seen at their locations is almost certainly associated with the lensing galaxies within the cluster. The angular resolution of the data is also not sufficient to determine if there is any radio emission associated with the gravitational arc, or the other lensed galaxies in the background of the cluster.

Assuming that SDSS~J2222+2745 has a typical radio spectral index of $\alpha=-0.7$ \citep{calistro2017}, the rest-frame 144 MHz luminosity density is found to be $L_{\rm 144~MHz} = 2.65\pm0.27 \times10^{24}$~W~Hz$^{-1}$. This is equivalent to a star-formation rate of $73^{+34}_{-22}$~M$_{\odot}$~yr$^{-1}$, under the assumption that all of the radio emission is associated with star formation.

\subsection{Lensed image flux ratios}

As gravitational lensing conserves surface brightness, the observed flux-ratios of the different lensed images can directly trace the relative magnifications provided by the lens. Any deviations from what is expected from the best lens model can therefore be used to infer if there is additional unaccounted for mass structure. This could be due to low mass dark matter haloes in the main lensing halo or along the line-of-sight to the background quasar, the abundance of which can directly test models for dark matter (e.g. \citealt{hsueh2020}). The mass distributions of massive clusters can also be quite complex, but sophisticated lens modelling has recently been used to test the cold dark matter model of galaxy formation with this method \citep{meneghetti2020}. At optical wavelengths, the observed flux-ratios can be changed (as a function of time) by microlensing, as quasar accretion disks are sufficiently small that lensing by stars in the lensing halo can occur. At radio wavelengths, source sizes are expected to be too large for microlensing to have a strong effect, and therefore, radio flux-ratios are best used to test the smooth lens mass model.

For SDSS~J1004+4112, we find that the flux ratio of lensed images A and B at 144 MHz is $F_{B/A} = 0.73\pm0.25$, which is in good agreement with the prediction of 0.66 from the lens model of \citet{oguri2010} and the measurement at 5 GHz of $F_{B/A} = 0.61\pm0.15$ by \citet{jackson2011}. For SDSS~J2222+2745, the 144 MHz flux-ratios between the three detected lensed images are $F_{B/A} = 0.81\pm0.18$ and $F_{C/A} = 0.66\pm0.17$. However, although the flux-ratio of lensed images A and B is in good agreement with the predicted magnification ratio of 0.74 from the lens model by \citet{sharon2017}, we find that the observed magnification of lensed image C is a factor of $1.5\pm0.4$ higher than expected; the predicted magnification-ratio is 0.46. Further observations will be needed to determine if this difference between the observed and predicted flux-ratios for image C of SDSS~J2222+2745 is significant and whether they change as a function of time.

\section{Nature of the radio emission} 
\label{discussion}

The underlying physical processes that lead to faint radio emission from extragalactic sources is not clear. It could be related to supernovae remnants or compact H{\sc ii} regions associated with massive star-formation (e.g. \citealt{mcdonald2002}), or from weak radio-jets (e.g. \citealt{radcliffe2018}), coronal emission (e.g. \citealt{laor2019}), thermal bremsstrahlung (e.g. \citealt{blundell2007}) or radiatively-driven shocks (e.g. \citealt{zakamska2014}) that are associated with AGN activity. There is also growing theoretical (e.g. \citealt{thomas2020}) and observational (e.g. \citealt{delvecchio2017,algera2020}) evidence that points towards a transition at $\mu$Jy-levels (cm wavelengths) from AGN dominated sources to those that are mainly ($>75$~percent) driven by star-formation.

For those radio sources associated with a quasar, the star-formation properties can only be determined via their FIR luminosities. However, there is a poorly understood scatter and redshift evolution in the radio--infrared correlation, which could potentially be accounted for through variations in the stellar mass of the host galaxies \citep{gurkan2018,smith2020} and/or through inverse Compton losses at high redshift \citep{schleicher2016}. In addition, even in sub-$\mu$Jy-level radio sources where the radio and FIR luminosities are consistent with the radio--infrared correlation for star-forming galaxies, observations with cm-wavelength very long baseline interferometry (VLBI) has demonstrated that AGN activity, in the form of small-scale jets and a high brightness-temperature core, is the dominant source of the radio emission \citep{hartley2019}.

In the case of the two high redshift radio sources investigated here, there is clearly evidence of AGN activity in the form of the quasar emission seen at optical and infrared wavelengths. However, it is not clear from the data in hand whether the weak radio emission observed at low frequencies with LOFAR is related to AGN activity or star formation. The radio spectrum of SDSS~J1004+4112 between 144 MHz and 5 GHz is about 1 to $1.3\sigma$ flatter than is typically observed for  AGN and star-forming galaxies, respectively (e.g. \citealt{calistro2017}). If the radio spectrum is genuinely flatter, this could be due to either free-free absorption or synchrotron self-absorption attenuating the spectrum at low frequencies, or alternatively, could be due to the presence of an optically thick radio source that dominates at high frequencies. In the case of the former, differentiating between these two absorption mechanisms is challenging unless the shape of the optically thick part of the spectrum at very low frequencies can be modelled (e.g. \citealt{mckean2016}). For the latter, cm-wavelength VLBI observations could be used to determine if there is evidence of AGN activity. However, such observations wouldn't necessarily rule out that a significant amount of the radio emission observed here is due to star formation; for example, the broadband spectrum could show a typical steep radio spectral index that is consistent with star formation at low frequencies, which flattens (or is inverted) at higher frequencies where an AGN component could dominate. Such a flattening in the radio spectra of AGN dominated sources from $\alpha_{\rm 150~MHz}^{\rm 1.4~GHz} = -0.66$ to $\alpha_{\rm 325~MHz}^{\rm 1.4~GHz} = -0.52$ has previously been reported \citep{calistro2017}. Further multi-frequency measurements between 150~MHz and 5~GHz are needed to determine if SDSS~J1004+4112 has a flat, single power-law  spectral energy distribution or not.

We note that SDSS~J1004+4112 (image A) is also included in the DR14 quasar sample of the SDSS, with a lensing-corrected absolute $i$-band magnitude of $M_i = -22.73\pm0.04$ \citep{myers2015}; this corresponds to a rest-frame luminosity-density of $L_{i~{\rm band}} = 5.47\pm0.20\times10^{22}$ W~Hz$^{-1}$. Using this and the radio luminosity-density determined above, we find that SDSS~J1004+4112 has a radio-loudness parameter of $R = 2.9\pm0.2$, which would place this object at the high-end of the distribution for optically-selected quasars at $1<z<2$ (see \citealt{gurkan2019} for details). This suggests that some fraction of the weak radio emission may be due to jets that are associated with AGN activity. However, due to micro-lensing by stars or compact objects within the lensing cluster \citep{mediavilla2009} and the intrinsic optical-variability of the quasar continuum emission \citep{fohlmeister2008}, the $r$-band magnitudes from lensed images A and B changed by $\Delta m = 0.75$ between 2003 and 2007. A decrease in the $i$-band emission at this level in 2015 would give $R \sim 1.5$. This would be in good agreement with the mean value of $R$ for the sample of optically-selected quasars at this epoch, whose radio emission is thought to be dominated by star formation \citep{gurkan2019}.

In the event that the radio emission observed here is due to star formation within the two quasar host galaxies, then their low star-formation rates, particularly in the case of SDSS~J1004+4112, could point toward systems at the end of their starburst phase, as predicted from galaxy formation models (e.g. \citealt{sanders1988}). To confirm this, further observations at mm-wavelengths would be needed to determine if there is a paucity of heated dust emission from obscured star formation, and to determine if there are disturbed gas kinematics or morphologies due to radiative feedback from the AGN or recent mergers (e.g. \citealt{paraficz2018,spingola2020}). Also, low molecular-gas fractions have been taken as evidence for the post-starburst phase in quasars \citep{stacey2020} and dusty star-forming galaxies \citep{spilker2016}. Therefore, further follow-up observations of low excitation CO could be used to investigate the evolutionary stage of SDSS~J1004+4112 and SDSS~J2222+2745.

The LOFAR imaging presented here can not unambiguously determine the nature of the radio emission from SDSS~J1004+4112 and SDSS~J2222+2745. However, we have demonstrated that low-luminosity radio emission is detectable from the population of sources that are expected to dominate the faint low-frequency sky at high redshifts. Such objects will be further investigated in larger numbers when the forthcoming LOFAR deep surveys are completed (e.g. \citealt{kondapally2020,sabater2020}). In the meantime, by combining gravitational lensing and LOFAR, it will be possible to identify objects like SDSS~J1004+4112 and SDSS~J2222+2745 for detailed multi-wavelength studies, so that we can better understand the composition of the faint radio source population at high redshift.

\section*{Acknowledgements}

JPM acknowledges support from the Netherlands Organization for Scientific Research (NWO, project number 629.001.023) and the Chinese Academy of Sciences (CAS, project number 114A11KYSB20170054). AD acknowledges support by the BMBF Verbundforschung under the grant 05A17STA. IP acknowledges support from INAF under the SKA/CTA PRIN ``FORECaST" and the PRIN MAIN STREAM ``SAuROS" projects. LOFAR \citep{vanhaarlem2013} is the Low Frequency Array designed and constructed by ASTRON. It has observing, data processing, and data storage facilities in several countries, which are owned by various parties (each with their own funding sources), and that are collectively operated by the ILT foundation under a joint scientific policy. The ILT resources have benefited from the following recent major funding sources: CNRS-INSU, Observatoire de Paris and Universit\'e d'Orl\'eans, France; BMBF, MIWF-NRW, MPG, Germany; Science Foundation Ireland (SFI), Department of Business, Enterprise and Innovation (DBEI), Ireland; NWO, The Netherlands; The Science and Technology Facilities Council, UK; Ministry of Science and Higher Education, Poland; The Istituto Nazionale di Astrofisica (INAF), Italy. The J\"ulich LOFAR Long Term Archive and the German LOFAR network are both coordinated and operated by the J\"ulich Supercomputing Centre (JSC), and computing resources on the supercomputer JUWELS at JSC were provided by the Gauss Centre for Supercomputing e.V. (grant CHTB00) through the John von Neumann Institute for Computing (NIC).

\section*{Data Availability}

The data used are available via the LOFAR and VLA archives, or on reasonable request to the corresponding author.



\bibliographystyle{mnras}
\bibliography{lofar-lens1} 

\begin{thebibliography}{}
\makeatletter
\relax
\def\mn@urlcharsother{\let\do\@makeother \do\$\do\&\do\#\do\^\do\_\do\%\do\~}
\def\mn@doi{\begingroup\mn@urlcharsother \@ifnextchar [ {\mn@doi@}
  {\mn@doi@[]}}
\def\mn@doi@[#1]#2{\def\@tempa{#1}\ifx\@tempa\@empty \href
  {http://dx.doi.org/#2} {doi:#2}\else \href {http://dx.doi.org/#2} {#1}\fi
  \endgroup}
\def\mn@eprint#1#2{\mn@eprint@#1:#2::\@nil}
\def\mn@eprint@arXiv#1{\href {http://arxiv.org/abs/#1} {{\tt arXiv:#1}}}
\def\mn@eprint@dblp#1{\href {http://dblp.uni-trier.de/rec/bibtex/#1.xml}
  {dblp:#1}}
\def\mn@eprint@#1:#2:#3:#4\@nil{\def\@tempa {#1}\def\@tempb {#2}\def\@tempc
  {#3}\ifx \@tempc \@empty \let \@tempc \@tempb \let \@tempb \@tempa \fi \ifx
  \@tempb \@empty \def\@tempb {arXiv}\fi \@ifundefined
  {mn@eprint@\@tempb}{\@tempb:\@tempc}{\expandafter \expandafter \csname
  mn@eprint@\@tempb\endcsname \expandafter{\@tempc}}}

\bibitem[\protect\citeauthoryear{{Algera} et~al.,}{{Algera}
  et~al.}{2020}]{algera2020}
{Algera} H.~S.~B.,  et~al., 2020, \mn@doi [\apj] {10.3847/1538-4357/abb77a},
  \href {https://ui.adsabs.harvard.edu/abs/2020ApJ...903..139A} {903, 139}

\bibitem[\protect\citeauthoryear{{Banerji}, {Carilli}, {Jones}, {Wagg},
  {McMahon}, {Hewett}, {Alaghband-Zadeh}  \& {Feruglio}}{{Banerji}
  et~al.}{2017}]{banerji2017}
{Banerji} M.,  {Carilli} C.~L.,  {Jones} G.,  {Wagg} J.,  {McMahon} R.~G.,
  {Hewett} P.~C.,  {Alaghband-Zadeh} S.,   {Feruglio} C.,  2017, \mn@doi
  [\mnras] {10.1093/mnras/stw3019}, \href
  {https://ui.adsabs.harvard.edu/abs/2017MNRAS.465.4390B} {465, 4390}

\bibitem[\protect\citeauthoryear{{Beelen}, {Cox}, {Benford}, {Dowell},
  {Kov{\'a}cs}, {Bertoldi}, {Omont}  \& {Carilli}}{{Beelen}
  et~al.}{2006}]{beelen2006}
{Beelen} A.,  {Cox} P.,  {Benford} D.~J.,  {Dowell} C.~D.,  {Kov{\'a}cs} A.,
  {Bertoldi} F.,  {Omont} A.,   {Carilli} C.~L.,  2006, \mn@doi [\apj]
  {10.1086/500636}, \href
  {https://ui.adsabs.harvard.edu/abs/2006ApJ...642..694B} {642, 694}

\bibitem[\protect\citeauthoryear{{Blundell} \& {Kuncic}}{{Blundell} \&
  {Kuncic}}{2007}]{blundell2007}
{Blundell} K.~M.,  {Kuncic} Z.,  2007, \mn@doi [\apjl] {10.1086/522695}, \href
  {https://ui.adsabs.harvard.edu/abs/2007ApJ...668L.103B} {668, L103}

\bibitem[\protect\citeauthoryear{{Calistro Rivera} et~al.,}{{Calistro Rivera}
  et~al.}{2017}]{calistro2017}
{Calistro Rivera} G.,  et~al., 2017, \mn@doi [\mnras] {10.1093/mnras/stx1040},
  \href {https://ui.adsabs.harvard.edu/abs/2017MNRAS.469.3468C} {469, 3468}

\bibitem[\protect\citeauthoryear{{Coppin} et~al.,}{{Coppin}
  et~al.}{2008}]{coppin2008}
{Coppin} K.~E.~K.,  et~al., 2008, \mn@doi [\mnras]
  {10.1111/j.1365-2966.2008.13553.x}, \href
  {https://ui.adsabs.harvard.edu/abs/2008MNRAS.389...45C} {389, 45}

\bibitem[\protect\citeauthoryear{{Dahle} et~al.,}{{Dahle}
  et~al.}{2013}]{dahle2013}
{Dahle} H.,  et~al., 2013, \mn@doi [\apj] {10.1088/0004-637X/773/2/146}, \href
  {https://ui.adsabs.harvard.edu/abs/2013ApJ...773..146D} {773, 146}

\bibitem[\protect\citeauthoryear{{Delvecchio} et~al.,}{{Delvecchio}
  et~al.}{2017}]{delvecchio2017}
{Delvecchio} I.,  et~al., 2017, \mn@doi [\aap] {10.1051/0004-6361/201629367},
  \href {https://ui.adsabs.harvard.edu/abs/2017A&A...602A...3D} {602, A3}

\bibitem[\protect\citeauthoryear{{Di Matteo}, {Springel}  \& {Hernquist}}{{Di
  Matteo} et~al.}{2005}]{dimateo2005}
{Di Matteo} T.,  {Springel} V.,   {Hernquist} L.,  2005, \mn@doi [\nat]
  {10.1038/nature03335}, \href
  {https://ui.adsabs.harvard.edu/abs/2005Natur.433..604D} {433, 604}

\bibitem[\protect\citeauthoryear{{Fohlmeister}, {Kochanek}, {Falco}, {Morgan}
  \& {Wambsganss}}{{Fohlmeister} et~al.}{2008}]{fohlmeister2008}
{Fohlmeister} J.,  {Kochanek} C.~S.,  {Falco} E.~E.,  {Morgan} C.~W.,
  {Wambsganss} J.,  2008, \mn@doi [\apj] {10.1086/528789}, \href
  {https://ui.adsabs.harvard.edu/abs/2008ApJ...676..761F} {676, 761}

\bibitem[\protect\citeauthoryear{{G{\"u}rkan} et~al.,}{{G{\"u}rkan}
  et~al.}{2018}]{gurkan2018}
{G{\"u}rkan} G.,  et~al., 2018, \mn@doi [\mnras] {10.1093/mnras/sty016}, \href
  {https://ui.adsabs.harvard.edu/abs/2018MNRAS.475.3010G} {475, 3010}

\bibitem[\protect\citeauthoryear{{G{\"u}rkan} et~al.,}{{G{\"u}rkan}
  et~al.}{2019}]{gurkan2019}
{G{\"u}rkan} G.,  et~al., 2019, \mn@doi [\aap] {10.1051/0004-6361/201833892},
  \href {https://ui.adsabs.harvard.edu/abs/2019A&A...622A..11G} {622, A11}

\bibitem[\protect\citeauthoryear{{Hartley}, {Jackson}, {Sluse}, {Stacey}  \&
  {Vives-Arias}}{{Hartley} et~al.}{2019}]{hartley2019}
{Hartley} P.,  {Jackson} N.,  {Sluse} D.,  {Stacey} H.~R.,   {Vives-Arias} H.,
  2019, \mn@doi [\mnras] {10.1093/mnras/stz510}, \href
  {https://ui.adsabs.harvard.edu/abs/2019MNRAS.485.3009H} {485, 3009}

\bibitem[\protect\citeauthoryear{{Heckman} \& {Best}}{{Heckman} \&
  {Best}}{2014}]{heckman2014}
{Heckman} T.~M.,  {Best} P.~N.,  2014, \mn@doi [\araa]
  {10.1146/annurev-astro-081913-035722}, \href
  {https://ui.adsabs.harvard.edu/abs/2014ARA&A..52..589H} {52, 589}

\bibitem[\protect\citeauthoryear{{Hodge} \& {da Cunha}}{{Hodge} \& {da
  Cunha}}{2020}]{hodge2020}
{Hodge} J.~A.,  {da Cunha} E.,  2020, \mn@doi [Royal Society Open Science]
  {10.1098/rsos.200556}, \href
  {https://ui.adsabs.harvard.edu/abs/2020RSOS....700556H} {7, 200556}

\bibitem[\protect\citeauthoryear{{Hopkins}, {Hernquist}, {Martini}, {Cox},
  {Robertson}, {Di Matteo}  \& {Springel}}{{Hopkins}
  et~al.}{2005}]{hopkins2005}
{Hopkins} P.~F.,  {Hernquist} L.,  {Martini} P.,  {Cox} T.~J.,  {Robertson} B.,
   {Di Matteo} T.,   {Springel} V.,  2005, \mn@doi [\apjl] {10.1086/431146},
  \href {https://ui.adsabs.harvard.edu/abs/2005ApJ...625L..71H} {625, L71}

\bibitem[\protect\citeauthoryear{{Hsueh}, {Enzi}, {Vegetti}, {Auger},
  {Fassnacht}, {Despali}, {Koopmans}  \& {McKean}}{{Hsueh}
  et~al.}{2020}]{hsueh2020}
{Hsueh} J.~W.,  {Enzi} W.,  {Vegetti} S.,  {Auger} M.~W.,  {Fassnacht} C.~D.,
  {Despali} G.,  {Koopmans} L.~V.~E.,   {McKean} J.~P.,  2020, \mn@doi [\mnras]
  {10.1093/mnras/stz3177}, \href
  {https://ui.adsabs.harvard.edu/abs/2020MNRAS.492.3047H} {492, 3047}

\bibitem[\protect\citeauthoryear{{Inada} et~al.,}{{Inada}
  et~al.}{2003}]{inada2003}
{Inada} N.,  et~al., 2003, \mn@doi [\nat] {10.1038/nature02153}, \href
  {https://ui.adsabs.harvard.edu/abs/2003Natur.426..810I} {426, 810}

\bibitem[\protect\citeauthoryear{{Jackson}}{{Jackson}}{2011}]{jackson2011}
{Jackson} N.,  2011, \mn@doi [\apjl] {10.1088/2041-8205/739/1/L28}, \href
  {https://ui.adsabs.harvard.edu/abs/2011ApJ...739L..28J} {739, L28}

\bibitem[\protect\citeauthoryear{{Kondapally} et~al.,}{{Kondapally}
  et~al.}{2021}]{kondapally2020}
{Kondapally} R.,  et~al., 2021, arXiv e-prints, \href
  {https://ui.adsabs.harvard.edu/abs/2020arXiv201108201K} {p. arXiv:2011.08201}

\bibitem[\protect\citeauthoryear{{Laor}, {Baldi}  \& {Behar}}{{Laor}
  et~al.}{2019}]{laor2019}
{Laor} A.,  {Baldi} R.~D.,   {Behar} E.,  2019, \mn@doi [\mnras]
  {10.1093/mnras/sty3098}, \href
  {https://ui.adsabs.harvard.edu/abs/2019MNRAS.482.5513L} {482, 5513}

\bibitem[\protect\citeauthoryear{{Madau} \& {Dickinson}}{{Madau} \&
  {Dickinson}}{2014}]{madau2014}
{Madau} P.,  {Dickinson} M.,  2014, \mn@doi [\araa]
  {10.1146/annurev-astro-081811-125615}, \href
  {https://ui.adsabs.harvard.edu/abs/2014ARA&A..52..415M} {52, 415}

\bibitem[\protect\citeauthoryear{{McDonald}, {Muxlow}, {Wills}, {Pedlar}  \&
  {Beswick}}{{McDonald} et~al.}{2002}]{mcdonald2002}
{McDonald} A.~R.,  {Muxlow} T.~W.~B.,  {Wills} K.~A.,  {Pedlar} A.,   {Beswick}
  R.~J.,  2002, \mn@doi [\mnras] {10.1046/j.1365-8711.2002.05580.x}, \href
  {https://ui.adsabs.harvard.edu/abs/2002MNRAS.334..912M} {334, 912}

\bibitem[\protect\citeauthoryear{{McKean} et~al.,}{{McKean}
  et~al.}{2016}]{mckean2016}
{McKean} J.~P.,  et~al., 2016, \mn@doi [\mnras] {10.1093/mnras/stw2105}, \href
  {https://ui.adsabs.harvard.edu/abs/2016MNRAS.463.3143M} {463, 3143}

\bibitem[\protect\citeauthoryear{{Mediavilla} et~al.,}{{Mediavilla}
  et~al.}{2009}]{mediavilla2009}
{Mediavilla} E.,  et~al., 2009, \mn@doi [\apj] {10.1088/0004-637X/706/2/1451},
  \href {https://ui.adsabs.harvard.edu/abs/2009ApJ...706.1451M} {706, 1451}

\bibitem[\protect\citeauthoryear{{Meneghetti} et~al.,}{{Meneghetti}
  et~al.}{2020}]{meneghetti2020}
{Meneghetti} M.,  et~al., 2020, \mn@doi [Science] {10.1126/science.aax5164},
  \href {https://ui.adsabs.harvard.edu/abs/2020Sci...369.1347M} {369, 1347}

\bibitem[\protect\citeauthoryear{{Myers} et~al.,}{{Myers}
  et~al.}{2015}]{myers2015}
{Myers} A.~D.,  et~al., 2015, \mn@doi [\apjs] {10.1088/0067-0049/221/2/27},
  \href {https://ui.adsabs.harvard.edu/abs/2015ApJS..221...27M} {221, 27}

\bibitem[\protect\citeauthoryear{{Nguyen}, {Lira}, {Trakhtenbrot}, {Netzer},
  {Cicone}, {Maiolino}  \& {Shemmer}}{{Nguyen} et~al.}{2020}]{nguyen2020}
{Nguyen} N.~H.,  {Lira} P.,  {Trakhtenbrot} B.,  {Netzer} H.,  {Cicone} C.,
  {Maiolino} R.,   {Shemmer} O.,  2020, \mn@doi [\apj]
  {10.3847/1538-4357/ab8bd3}, \href
  {https://ui.adsabs.harvard.edu/abs/2020ApJ...895...74N} {895, 74}

\bibitem[\protect\citeauthoryear{{Oguri}}{{Oguri}}{2010}]{oguri2010}
{Oguri} M.,  2010, \mn@doi [\pasj] {10.1093/pasj/62.4.1017}, \href
  {https://ui.adsabs.harvard.edu/abs/2010PASJ...62.1017O} {62, 1017}

\bibitem[\protect\citeauthoryear{{Paraficz} et~al.,}{{Paraficz}
  et~al.}{2018}]{paraficz2018}
{Paraficz} D.,  et~al., 2018, \mn@doi [\aap] {10.1051/0004-6361/201731250},
  \href {https://ui.adsabs.harvard.edu/abs/2018A&A...613A..34P} {613, A34}

\bibitem[\protect\citeauthoryear{{Planck Collaboration} et~al.,}{{Planck
  Collaboration} et~al.}{2016}]{planck2016}
{Planck Collaboration} et~al., 2016, \mn@doi [\aap]
  {10.1051/0004-6361/201525830}, \href
  {https://ui.adsabs.harvard.edu/abs/2016A&A...594A..13P} {594, A13}

\bibitem[\protect\citeauthoryear{{Radcliffe} et~al.,}{{Radcliffe}
  et~al.}{2018}]{radcliffe2018}
{Radcliffe} J.~F.,  et~al., 2018, \mn@doi [\aap] {10.1051/0004-6361/201833399},
  \href {https://ui.adsabs.harvard.edu/abs/2018A&A...619A..48R} {619, A48}

\bibitem[\protect\citeauthoryear{{Sabater} et~al.,}{{Sabater}
  et~al.}{2021}]{sabater2020}
{Sabater} J.,  et~al., 2021, arXiv e-prints, \href
  {https://ui.adsabs.harvard.edu/abs/2020arXiv201108211S} {p. arXiv:2011.08211}

\bibitem[\protect\citeauthoryear{{Sanders}, {Soifer}, {Elias}, {Madore},
  {Matthews}, {Neugebauer}  \& {Scoville}}{{Sanders}
  et~al.}{1988}]{sanders1988}
{Sanders} D.~B.,  {Soifer} B.~T.,  {Elias} J.~H.,  {Madore} B.~F.,  {Matthews}
  K.,  {Neugebauer} G.,   {Scoville} N.~Z.,  1988, \mn@doi [\apj]
  {10.1086/165983}, \href
  {https://ui.adsabs.harvard.edu/abs/1988ApJ...325...74S} {325, 74}

\bibitem[\protect\citeauthoryear{{Schaye} et~al.,}{{Schaye}
  et~al.}{2015}]{schaye2015}
{Schaye} J.,  et~al., 2015, \mn@doi [\mnras] {10.1093/mnras/stu2058}, \href
  {https://ui.adsabs.harvard.edu/abs/2015MNRAS.446..521S} {446, 521}

\bibitem[\protect\citeauthoryear{{Schleicher} \& {Beck}}{{Schleicher} \&
  {Beck}}{2016}]{schleicher2016}
{Schleicher} D. R.~G.,  {Beck} R.,  2016, \mn@doi [\aap]
  {10.1051/0004-6361/201628843}, \href
  {https://ui.adsabs.harvard.edu/abs/2016A&A...593A..77S} {593, A77}

\bibitem[\protect\citeauthoryear{{Sharon} et~al.,}{{Sharon}
  et~al.}{2017}]{sharon2017}
{Sharon} K.,  et~al., 2017, \mn@doi [\apj] {10.3847/1538-4357/835/1/5}, \href
  {https://ui.adsabs.harvard.edu/abs/2017ApJ...835....5S} {835, 5}

\bibitem[\protect\citeauthoryear{{Shimwell} et~al.,}{{Shimwell}
  et~al.}{2019}]{shimwell2019}
{Shimwell} T.~W.,  et~al., 2019, \mn@doi [\aap] {10.1051/0004-6361/201833559},
  \href {https://ui.adsabs.harvard.edu/abs/2019A&A...622A...1S} {622, A1}

\bibitem[\protect\citeauthoryear{{Smith}, {Koss}, {Mushotzky}, {Wong},
  {Shimizu}, {Ricci}  \& {Ricci}}{{Smith} et~al.}{2020}]{smith2020}
{Smith} K.~L.,  {Koss} M.,  {Mushotzky} R.,  {Wong} O.~I.,  {Shimizu} T.~T.,
  {Ricci} C.,   {Ricci} F.,  2020, \mn@doi [\apj] {10.3847/1538-4357/abc3c4},
  \href {https://ui.adsabs.harvard.edu/abs/2020ApJ...904...83S} {904, 83}

\bibitem[\protect\citeauthoryear{{Spilker}, {Bezanson}, {Marrone}, {Weiner},
  {Whitaker}  \& {Williams}}{{Spilker} et~al.}{2016}]{spilker2016}
{Spilker} J.~S.,  {Bezanson} R.,  {Marrone} D.~P.,  {Weiner} B.~J.,  {Whitaker}
  K.~E.,   {Williams} C.~C.,  2016, \mn@doi [\apj]
  {10.3847/0004-637X/832/1/19}, \href
  {https://ui.adsabs.harvard.edu/abs/2016ApJ...832...19S} {832, 19}

\bibitem[\protect\citeauthoryear{{Spingola} et~al.,}{{Spingola}
  et~al.}{2020}]{spingola2020}
{Spingola} C.,  et~al., 2020, \mn@doi [\mnras] {10.1093/mnras/staa1342}, \href
  {https://ui.adsabs.harvard.edu/abs/2020MNRAS.495.2387S} {495, 2387}

\bibitem[\protect\citeauthoryear{{Stacey} et~al.,}{{Stacey}
  et~al.}{2018}]{stacey2018}
{Stacey} H.~R.,  et~al., 2018, \mn@doi [\mnras] {10.1093/mnras/sty458}, \href
  {https://ui.adsabs.harvard.edu/abs/2018MNRAS.476.5075S} {476, 5075}

\bibitem[\protect\citeauthoryear{{Stacey} et~al.,}{{Stacey}
  et~al.}{2019}]{stacey2019}
{Stacey} H.~R.,  et~al., 2019, \mn@doi [\aap] {10.1051/0004-6361/201833967},
  \href {https://ui.adsabs.harvard.edu/abs/2019A&A...622A..18S} {622, A18}

\bibitem[\protect\citeauthoryear{{Stacey} et~al.,}{{Stacey}
  et~al.}{2021}]{stacey2020}
{Stacey} H.~R.,  et~al., 2021, \mn@doi [\mnras] {10.1093/mnras/staa3433}, \href
  {https://ui.adsabs.harvard.edu/abs/2021MNRAS.500.3667S} {500, 3667}

\bibitem[\protect\citeauthoryear{{Tasse} et~al.,}{{Tasse}
  et~al.}{2021}]{tasse2020}
{Tasse} C.,  et~al., 2021, arXiv e-prints, \href
  {https://ui.adsabs.harvard.edu/abs/2020arXiv201108328T} {p. arXiv:2011.08328}

\bibitem[\protect\citeauthoryear{{Thomas}, {Dav\'e}, {Jarvis}  \&
  {Angl\'es-Alc\'azar}}{{Thomas} et~al.}{2021}]{thomas2020}
{Thomas} N.,  {Dav\'e} D.,  {Jarvis} M.~J.,   {Angl\'es-Alc\'azar} D.,  2021,
  arXiv e-prints, \href {https://ui.adsabs.harvard.edu/abs/2020arXiv200901277S}
  {p. arXiv:2010.11225}

\bibitem[\protect\citeauthoryear{{Vogelsberger} et~al.,}{{Vogelsberger}
  et~al.}{2014}]{vogelsberger2014}
{Vogelsberger} M.,  et~al., 2014, \mn@doi [\nat] {10.1038/nature13316}, \href
  {https://ui.adsabs.harvard.edu/abs/2014Natur.509..177V} {509, 177}

\bibitem[\protect\citeauthoryear{{Wang} et~al.,}{{Wang}
  et~al.}{2019}]{wang2019}
{Wang} L.,  et~al., 2019, \mn@doi [\aap] {10.1051/0004-6361/201935913}, \href
  {https://ui.adsabs.harvard.edu/abs/2019A&A...631A.109W} {631, A109}

\bibitem[\protect\citeauthoryear{{Zakamska} \& {Greene}}{{Zakamska} \&
  {Greene}}{2014}]{zakamska2014}
{Zakamska} N.~L.,  {Greene} J.~E.,  2014, \mn@doi [\mnras]
  {10.1093/mnras/stu842}, \href
  {https://ui.adsabs.harvard.edu/abs/2014MNRAS.442..784Z} {442, 784}

\bibitem[\protect\citeauthoryear{{van Haarlem} et~al.,}{{van Haarlem}
  et~al.}{2013}]{vanhaarlem2013}
{van Haarlem} M.~P.,  et~al., 2013, \mn@doi [\aap]
  {10.1051/0004-6361/201220873}, \href
  {https://ui.adsabs.harvard.edu/abs/2013A&A...556A...2V} {556, A2}

\makeatother
\end{thebibliography}








\bsp	
\label{lastpage}
\end{document}